\documentclass[preprint,showkeys]{revtex4}
\usepackage{graphicx}
\usepackage{dcolumn}
\usepackage{amsmath}
\usepackage{bm}
\newcommand{\be}{\begin{equation}}
\newcommand{\ee}{\end{equation}}
\newcommand{\ba}{\begin{eqnarray}}
\newcommand{\ea}{\end{eqnarray}}
\begin{document}
\title{Thermodynamic and structural anomalies \\
of the Gaussian-core model in one dimension}
\author{Cristina Speranza~\footnote{Email: \tt cristina.speranza@unime.it}, Santi Prestipino~\footnote{Email: \tt santi.prestipino@unime.it}, and Paolo V. Giaquinta~\footnote{Email: {\tt paolo.giaquinta@unime.it} (corresponding author)}}
\affiliation{Universit\`a degli Studi di Messina, Dipartimento di Fisica,
Contrada Papardo, I-98166 Messina, Italy}
\begin{abstract}
We investigated the equilibrium properties of a one-dimensional system of classical particles which interact in pairs through a bounded repulsive potential with a Gaussian shape. Notwithstanding the absence of a proper fluid-solid phase transition, we found that the system exhibits a complex behaviour, with ``anomalies'' in the density and in the thermodynamic response functions which closely recall those observed in bulk and confined liquid water.  We also discuss the emergence in the cold fluid under compression of an unusual structural regime, characterized by density correlations reminiscent of the ordered arrangements found in clustered crystals. 
\end{abstract}
\maketitle
\section{Introduction}
Renewed interest has recently emerged in the thermodynamic and transport properties of one-dimensional (1D) systems, also in view of their potential relevance to the modelling of interacting colloidal particles diffusing within narrow channels~\cite{channels}. The spontaneous emergence in such systems of (even intense) local ordering phenomena, not necessarily accompanied by proper phase transitions, is a topical issue that continues to be investigated also in relation to its effects on the dynamics of the particles~\cite{velarde}. In this regard, a landmark result still is the celebrated van Hove's theorem~\cite{VanHove} according to which a system of identical particles with a hard core and pairwise interactions extending over a finite range has no phase transition in 1D, in the absence of external fields. The general issue of the existence (or non-existence) of phase transitions in 1D systems with short-ranged interactions has been critically reviewed in more recent years, among others, by Cuesta and S\'anchez~\cite{cuesta}. A way to potentially escape the negative verdict of van Hove's theorem is to allow for {\em bounded} repulsive interactions between particles. The simplest of such models is a system of penetrable spheres (PS), whose interaction potential takes a finite positive value whenever two particles overlap, while vanishing otherwise~\cite{marquest,santos1}. In the absence of an infinite repulsive barrier, interactions are no longer restricted to nearest neighbours. As a result, at variance with the Tonks gas~\cite{tonks}, no analytical solution of the PS model is currently known. Fantoni~\cite{Fantoni} has recently provided robust arguments against the existence of a phase transition in a 1D system of penetrable particles with a negative short-ranged square well outside the repulsive core~\cite{santos2}.

A largely studied variant of bounded repulsive interactions is the Gaussian potential~\cite{Stillinger}. Its continuing interest also stems from its frequent applications as a model interaction in soft-matter systems, such as the dilute solutions of highly ramified polymers in a good solvent (star polymers), where the ``effective'' repulsion between the centres of mass of two polymer chains can be typically described through a Gaussian law~\cite{likos}. The corresponding Gaussian-core model (GCM) can at most exhibit fluid-solid and, possibly, solid-solid phase transitions since a liquid-vapour transition is clearly excluded by the absence of an attractive component in the potential. In fact, at low enough temperature three-dimensional (3D) Gaussian particles crystallize, upon compression, in a cubic solid structure, with either a face-centred or a body-centred symmetry~\cite{Prestipino1}. However, a further increase of the density does eventually lead to a reentrant melting of the solid. This phenomenon has been observed also in two dimensions (2D) where, at variance with the 3D case, a narrow hexatic region appears between the isotropic fluid phase and the triangular solid phase all along the melting line~\cite{Prestipino2}.

As for the phase stability of the GCM in 1D, no rigorous proof of the absence of a phase transition has been presented so far. However, Fantoni conjectured that his arguments against the existence of a phase transition in a 1D system of attractive penetrable spheres apply equally well, under appropriate conditions, to a larger class of model fluids with a decaying long-ranged repulsive tail, including the GCM~\cite{Fantoni}.

Whether or not the GCM fluid crystallizes in 1D, one can safely expect a far from trivial phase behaviour because of the ever-increasing occurrence, under compression, of soft particle-core overlaps.  A natural candidate for the macrostate of minimum Gibbs free energy is a diffusely ordered arrangement formed by more or less equally spaced particles, as is the case of 1D hard rods at high density~\cite{pvg}. However, the thermodynamic competition between energy and entropy may also lead, in a fluid with a soft bounded potential, to even more complex arrangements such as those found in ``clustered'' or ``tower'' crystals, where two or more (superimposed) particles are confined within the same cell~\cite{likos}.
\section{Model and method}
We consider a system of point particles repelling each other, at a relative distance $r$, through a Gaussian pair potential
\be
u(r)=\epsilon\exp\left[-(r/\sigma)^2\right]\,,
\label{eq1}
\ee
where $\epsilon$ and $\sigma$ fix the energy and length scales, respectively. The GCM potential  has an inflection point at  $r_0=\sigma/\sqrt{2}$, where its curvature changes from concave to convex. Correspondingly, for $r\le r_0$ the strength of the force between two particles, $f(r)=-\mathrm{d} u(r) / \mathrm{d} r$, decreases as the particles approach each other. However, it is over a larger range $(0\le r \le \sigma)$ that the local virial function, $rf(r)$, and thus the contribution of a pair of interacting molecules to the pressure of the system, decreases when the separation also decreases. This condition is typically taken as the signature of ``core softening''~\cite{debenedetti}. As is well known, a core-softened -- in the just specified sense -- potential can generate a density anomaly, associated with  a negative thermal expansion coefficient. This circumstance was originally verified for the GCM fluid in 3D by Stillinger and Weber~\cite{sw}. Indeed, a number of thermodynamic, structural and dynamical properties of the GCM fluid have been already found to exhibit, both in 2D and 3D, ``waterlike'' anomalies which render the behaviour of this model qualitatively different from that of a ``simple'' ({\em i.e.}, Lennard-Jones-like) fluid.

To our knowledge, no systematic numerical study of the 1D GCM fluid has been undertaken so far. In this paper we present the results of an investigation carried out with the Monte Carlo (MC) method in the isothermal-isobaric ensemble. The data we present were obtained with samples of $N=200$ particles, unless otherwise specified; however, we also tested the stability and convergence of the results with larger samples of $500$ and $1000$ particles. No qualitatively significant changes emerged upon increasing the ``size'' of the calculation. We collected data over trajectories of, typically, five million sweeps, every sweep consisting of $N+1$ elementary MC moves, including one attempt, on average, to change the volume $V$ of the system. During the equilibration runs, we adjusted the maximum particle displacement and the volume change so as to maintain the acceptance rates of both types of moves close to $50\%$ in the production runs. We carried out our simulations along a number of isothermal and isobaric paths, from low to high density and from high to low temperatures, continuing at each state point from the last system configuration produced in the previous run. We computed, for given values of $T$ and $P$, the following properties: the average number density, $n=N/\langle V\rangle$; the average energy, $E=Nk_{\mathrm B}T/2+\langle U\rangle$, where $k_{\mathrm B}$ is Boltzmann's constant and $U$ is the total potential energy; the specific heat at constant pressure, $C_P=T(\partial s/\partial T)_P$, where $s$ is the entropy per particle; the isothermal compressibility, $K_T=-v^{-1}(\partial v/\partial P)_T$, where $v=n^{-1}$ is the average volume per particle; the thermal expansion coefficient, $\alpha_P=v^{-1}(\partial v/\partial T)_P$. The three thermodynamic response functions were obtained as thermal averages of covariances of the fluctuating variables (energy and volume): $C_P/k_{\mathrm B}=\langle \left[ \Delta(E+PV) \right]^2 \rangle/(k_{\mathrm B}T)^2$, $\alpha_T/k_{\mathrm B}=\langle\Delta(E+PV)\Delta V \rangle/\left[ \langle V \rangle (k_{\mathrm B}T)^2 \right]$, and $K_T=\langle (\Delta V)^2\rangle/\left[\langle V \rangle (k_{\mathrm B}T)\right]$, where $\Delta X=X-\langle X \rangle$.

 We also calculated the radial distribution function (RDF), $g(|r|)=v\langle\sum_{k\neq 1}\delta(x_k-x_1-r)\rangle$, and the associated structure factor, $S(|q|)=1+n\int_{-\infty}^{\infty} \mathrm{d}x \exp(-iqx) \left[ g(|x|)-1 \right]$.

The Monte Carlo study was complemented with the results obtained with the hypernetted-chain (HNC) approximation and with exact total-energy calculations carried out at $T=0$ with increasing pressure for several candidates for the solid phase, in order to gain insight into the preferred forms of particle aggregation at low temperatures.

In the following, we shall also make use of reduced units for density, temperature, and pressure: $\rho=n\sigma$, $\tau=k_{\mathrm B}T/\epsilon$, and $\Pi=P\sigma/\epsilon$.
\section{Results}
In the GCM fluid the growth of density correlations upon compression is eventually frustrated by the finite strength of the repulsion between particles and by the decreasing strength of their mutual force as they approach each other. We anticipate that for densities larger than the value corresponding to a nearest-neighbour (NN) distance roughly equal to $1.5\sigma~ (\Pi\simeq0.35)$, a further increase of the pressure causes a suppression, rather than a sharpening, of the local structure of the fluid since in a sufficiently dense environment the overlapping of two particles entails an entropy gain that is larger than the associated energy penalty. Lang and coworkers~\cite{Likos} coined the term ``infinite-density ideal gas'' to represent the uncorrelated $(g(r)=1)$ behaviour of Gaussian particles at very high densities.

\begin{figure}
\centerline{\includegraphics[width=0.7\textwidth]{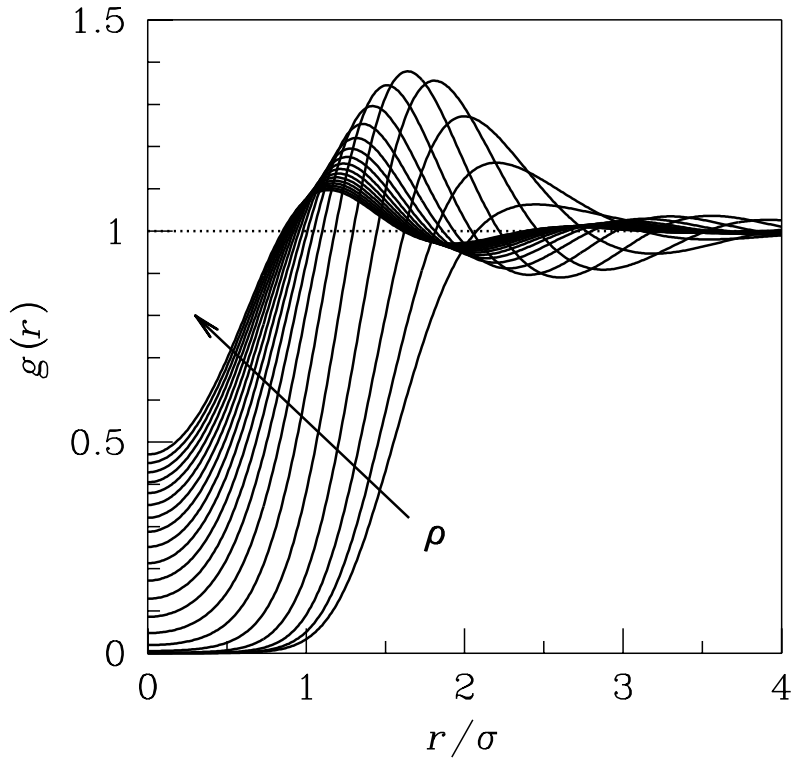}}
\caption{Radial distribution function in the HNC
approximation plotted as a function of the distance for a reduced temperature $\tau=0.1$ and for reduced densities $\rho=0.1,0.2,\ldots,1.9,2$ (the density increases along the direction indicated by the arrow).}
\label{HNC}
\end{figure}

Such a trend is already manifest in the RDF calculated through the HNC approximation at the reduced temperature $\tau=0.1$ (see Fig.~\ref{HNC}) which, though quantitatively accurate only for $\rho\gtrsim 2$, yields the correct qualitative trend of the local fluid structure with increasing pressure. We see from the picture that the NN distance, corresponding to the position of the first maximum in the RDF, steadily decreases upon compression but its statistical definition as a microscopic length scale is highest at an intermediate pressure corresponding to a reduced density $\rho\simeq 0.5$.
\begin{figure}
\centerline{\includegraphics[width=0.7\textwidth]{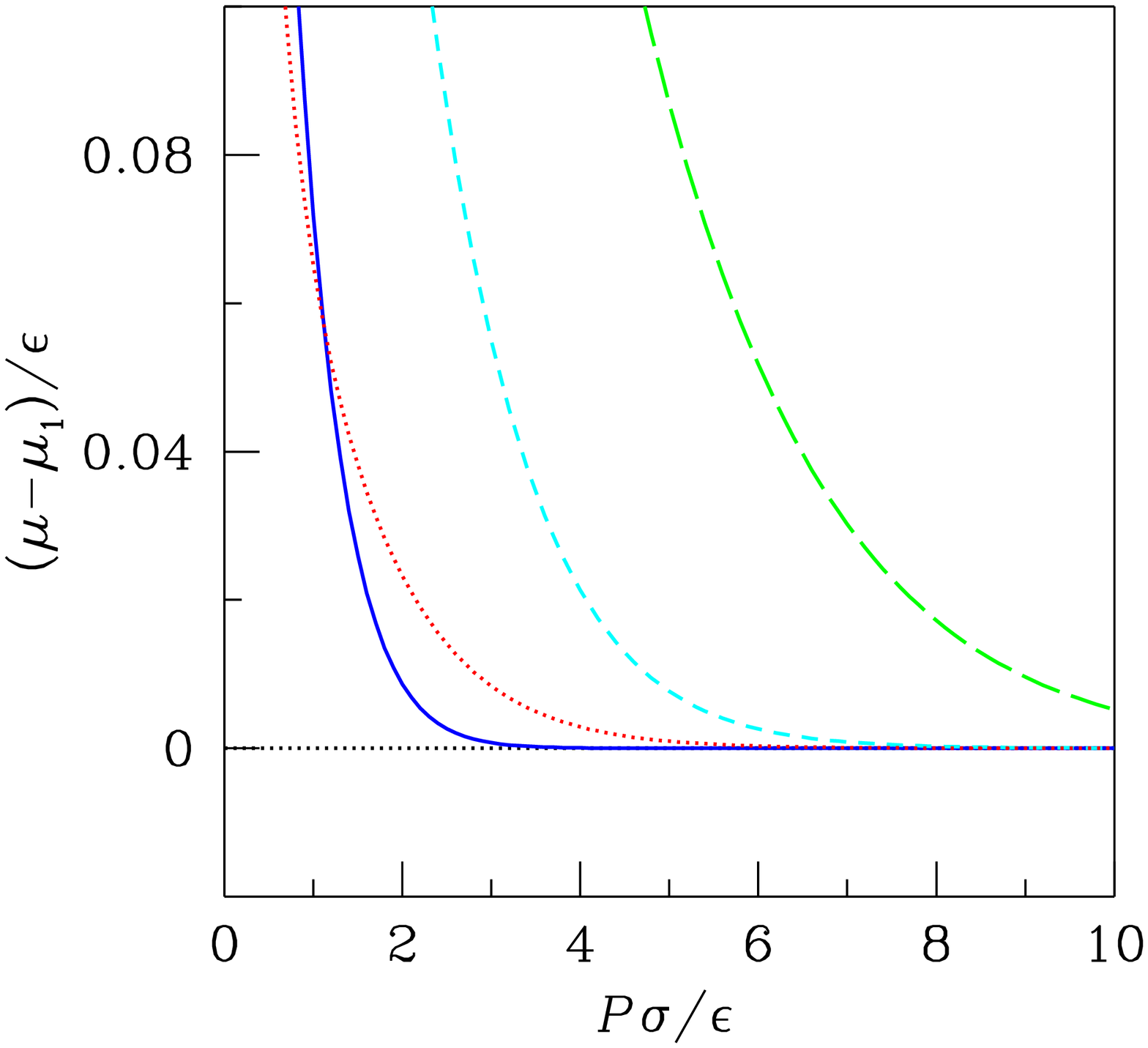}}
\caption{Excess chemical potential, relative to the value ($\mu_1$) in an ordered state of equally spaced particles, plotted as a function of pressure at $T=0$, for clustered crystalline phases: continuous blue line, 2-clustered crystal; dashed cyanide line, 3-clustered crystal; long-dashed green line, 4-clustered crystal. The dotted red line yields the excess chemical potential of the state formed by regularly alternating singles and pairs.}
\label{chemical_potential}
\end{figure}
\subsection{Phase stability at $T=0$}
Before examining the MC findings, it is worth considering the thermodynamic behaviour of the fluid at zero temperature. Stillinger proved that in the limit of low $T$ and $n$ the thermodynamics of the GCM fluid reduces, for any space dimensionality, to that of hard spheres with a temperature-dependent diameter $d(\beta)=\sigma\sqrt{\ln(\beta\epsilon)}$, where $\beta=(k_{\mathrm B}T)^{-1}$~\cite{Stillinger}. In one dimension, the hard-rod behaviour sets in for $\rho_{\mathrm{HS}}=\sigma/d$; below this threshold the equilibrium properties of the 1D GCM fluid can be mapped onto those of the Tonks gas. However, $\rho_{\mathrm{HS}}$ is too small a density (about $0.40$ for $\tau=0.002$, corresponding to a pressure slightly smaller than $0.05$) for the hard-sphere connection to be of any relevance to the present inquiry into the existence of phase transitions in the 1D GCM fluid at high pressure.

The competition for thermodynamic stability between single-occupancy and clustered solids is ruled by the chemical potential which, at zero temperature, is equal to the enthalpy per particle: $\mu=e+Pv$; moreover, any crystalline phase consists of just one microstate (modulo  a translation). We first consider an arrangement of equally spaced particles:
\be
\mu_{\mathrm1}=\min_{n}\left\{\sum_{m=1}^{\infty}\exp\left(-\frac{m^2}{n^2}\right)+
\frac{P}{n}\right\}\,.
\label{eq2}
\ee
We then compare $\mu_{\mathrm1}$ with the chemical potentials of clustered phases with equidistant lumps of $k=2,3,\ldots$ particles, all lying at the same positions:
\be 
\mu_k=\frac{k-1}{2}+\min_{n}\left\{k\sum_{m=1}^{\infty}
\exp\left(-\frac{k^2m^2}{n^2}\right)+\frac{P}{n}\right\}\,.
\label{eq3}
\ee
A further possibility that we have considered is a state in which isolated particles regularly alternate with pairs:
\be
\mu_{\mathrm(1,2)}=\frac{1}{3}+\min_{n}\left\{\frac{4}{3}\sum_{m=1,3,\ldots}
\exp\left(-\frac{9m^2}{4n^2}\right)+\frac{5}{3}\sum_{m=2,4,\ldots}
\exp\left(-\frac{9m^2}{4n^2}\right)+\frac{P}{n}\right\}\,.
\label{eq4}
\ee
In Fig.~\ref{chemical_potential} we plot all such chemical potentials as a function of the pressure. The comparison shows that the crystal formed by equidistant particles is the most stable phase at zero temperature, whatever the pressure. No better solution was obtained upon allowing for a small fixed separation between the particles forming a pair in either the $2-2-2-2-\ldots$ or the $1-2-1-2-\ldots$ arrangement. This result is a strong indication of the absence of any kind of phase transition in the 1D GCM fluid. However, the free-energy penalties associated with the nucleation of clustered solids progressively decrease with the pressure, the faster the smaller the number of particles in a given cell. Hence, one cannot  exclude {\em a priori} the possibility that, at non-zero temperature and high enough pressure, such phases may enter the thermodynamic game for entropic reasons. In passing, we note that for $\Pi\gtrsim 3$ the 2-clustered crystal is almost degenerate with the single-occupancy crystal.
\subsection{Thermodynamic properties}
\begin{figure}
\centerline{\includegraphics[width=0.7\textwidth]{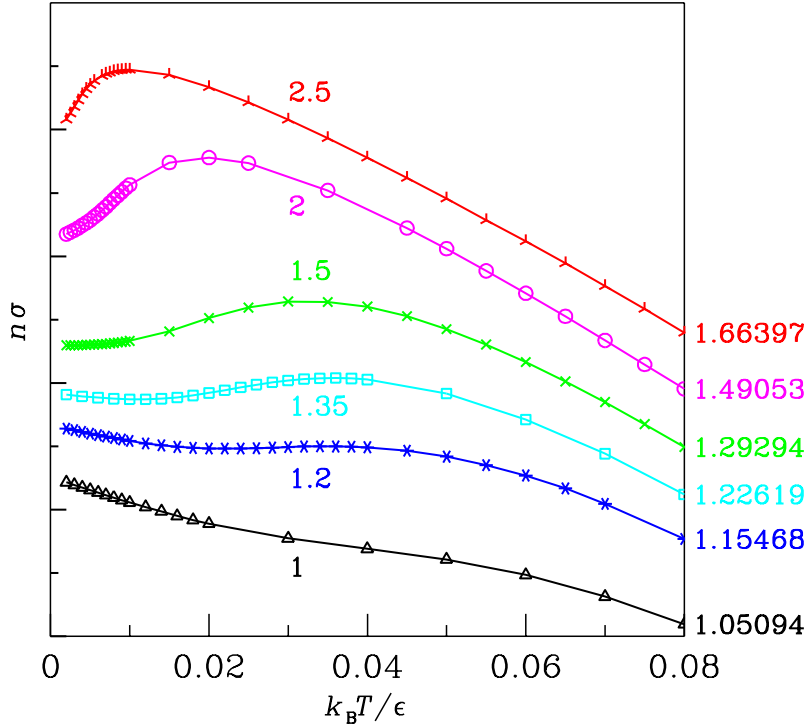}}
\caption{Reduced number density plotted as a function of the reduced temperature with increasing pressure: triangles (black), $\Pi=1$; stars (blue), $\Pi=1.2$; squares (cyanide), $\Pi=1.35$; crosses (green), $\Pi=1.5$; circles (magenta), $\Pi=2$; tripods (red), $\Pi=2.5$. To improve readability, the curves were shifted vertically by changing amounts; the separation between two major tick marks along the left vertical axis is equal to $0.01$, while the absolute scale can be fixed through the value of each curve at $\tau=0.08$, that is reported along the right vertical axis.}
\label{density}
\end{figure}
Following the computational procedure described in Sect. II, we investigated the phase behaviour of the 1D GCM fluid at reduced temperatures $0.002\le \tau\le1$ and for reduced pressures $0.05\le \Pi\le3.5$. We verified that neither the number density nor the energy exhibit any (even rounded-off) jump discontinuity over the explored domain.  We also performed a chain of isothermal simulations at high pressure starting from an ordered 2-clustered configuration but did not find any sign of hysteresis in either the density or the energy. Hence, on a strictly thermodynamic basis, the equilibrium state of the investigated model is always that of  a fluid. However, we found out that at low enough temperatures this fluid undergoes remarkable structural changes, with an anomalous ({\em i.e.}, nonstandard) thermodynamic behaviour which, in many respects, closely resembles that observed in liquid water, when gradually cooled from above the freezing temperature into the deeply metastable region (possibly in a confined environment). 

The signature of such a ``complex-fluid'' behaviour is the emergence of a volumetric anomaly in the dense fluid: as shown in Fig.~\ref{density}, at low pressure the density decreases monotonically with the temperature; however, for $\Pi \gtrsim 1.2$ the trend turns from decreasing to increasing over a limited temperature range $T_{\mathrm{min}}(P)\le T\le T_{\mathrm{max}}(P)$. Correspondingly, a maximum shows up in $n(T,P)$ at $T=T_{\mathrm{max}}(P)$ which, for $\Pi\lesssim1.5$, is preceded by a minimum at a nonzero temperature, $T_{\mathrm{min}}(P)$. For $\Pi\gtrsim1.5$, the density increases with $T$ from $T=0$, until it inverts its trend at the temperature of maximum density (TMD).

A maximum in the density of the GCM fluid had already been observed in higher space dimensions~\cite{sw,Prestipino2}: in this respect, it is altogether remarkable to verify the persistence of this feature in one dimension. But even more remarkable is the emergence of a shallow minimum in $n(T)$ over a range of pressures: the fluid behaves in an anomalous way for $T_{\mathrm{min}}(P)\le T\le T_{\mathrm{max}}(P)$ but turns ``normal'' again below the temperature of minimum density (TmD). In real liquids a density minimum has been detected in even fewer cases than those, already rare, where a maximum is observed. Two noteworthy examples are bulk tellurium~\cite{tscuchiya1}, also mixed with sulphur~\cite{tscuchiya2}, and (supercooled) confined water~\cite{liu,faraone,mallamace}, where the emergence of a minimum has been associated with the formation of a defect-free ``random tetrahedral network''~\cite{poole}. However, the profound difference between tellurium and water (as noted by Angell, tellurium is not known as a network liquid~\cite{angell}) makes it clear that even the TmD is caused by some more basic feature of the effective interaction potential.
\begin{figure}
\centerline{\includegraphics[width=0.75\textwidth]{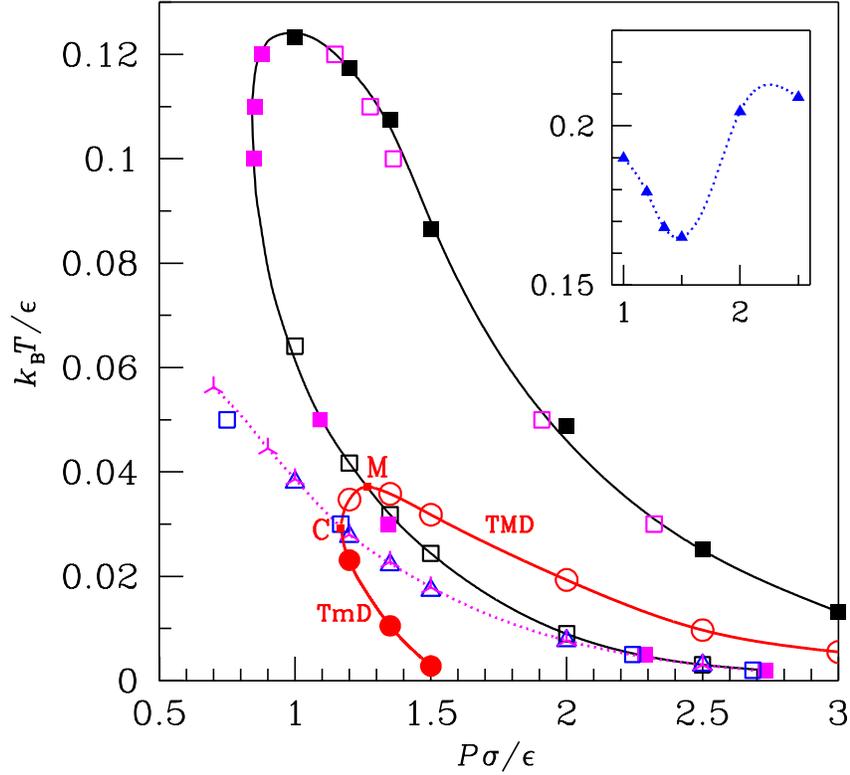}}
\caption{Location of the density and response functions extrema in the $(P,T)$ plane. Density: open red circles, temperature maxima (TMD); solid red circles, temperature minima (TmD). Coefficient of thermal expansion: magenta tripods, temperature minima; 
open magenta squares, pressure maxima; solid magenta squares, pressure minima. Isobaric specific heat: open blue triangles, temperature maxima; open blue squares, pressure maxima; solid blue triangles (inset), temperature minima. Isothermal compressibility: open black squares, temperature maxima; solid black squares, temperature minima. The locations of the symbols correspond to numerical estimates of the thermodynamic coordinates of the extrema obtained via Monte Carlo simulations along either isothermal or isobaric paths. The lines through the data are guides to the eye.}
\label{loci}
\end{figure}

The relevant thermodynamic loci corresponding to the location of the extrema of the density and of the response functions are plotted in Fig.~\ref{loci}. The volumetric anomaly region, corresponding to a negative value of the coefficient of thermal expansion, is the region bounded by a maximum and (possibly) a minimum of the density; all along this boundary the expansivity vanishes. The TMD and TmD lines are seen to originate from point C (with coordinates $\Pi_{\mathrm C}\simeq 1.169$ and $\tau_{\mathrm C}\simeq 0.029$), where the first and second isobaric temperature derivatives of the density both vanish $\left[\alpha_P=(\partial\alpha_P/\partial T)_{P=P_C}=0\right]$. This point marks the onset of the volumetric anomaly in the fluid and is apparently reached with infinite slope along the TMD and TmD lines. The TMD line is seen to pass through a maximum (M), with coordinates $\Pi_{\mathrm{M}}\simeq 1.267$ and $\tau_{\mathrm{M}}\simeq 0.037$; correspondingly, the TMD locus has a positive slope for $P_{\mathrm C} \le P < P_{\mathrm{M}}$, while decreasing asymptotically to zero with increasing pressure for $P > P_{\mathrm{M}}$. Instead, the TmD shows a rapidly decreasing monotonic trend on compression and appears to vanish at a {\em finite} reduced pressure just larger than $1.5$.
\begin{figure}
\centerline{\includegraphics[width=0.75\textwidth]{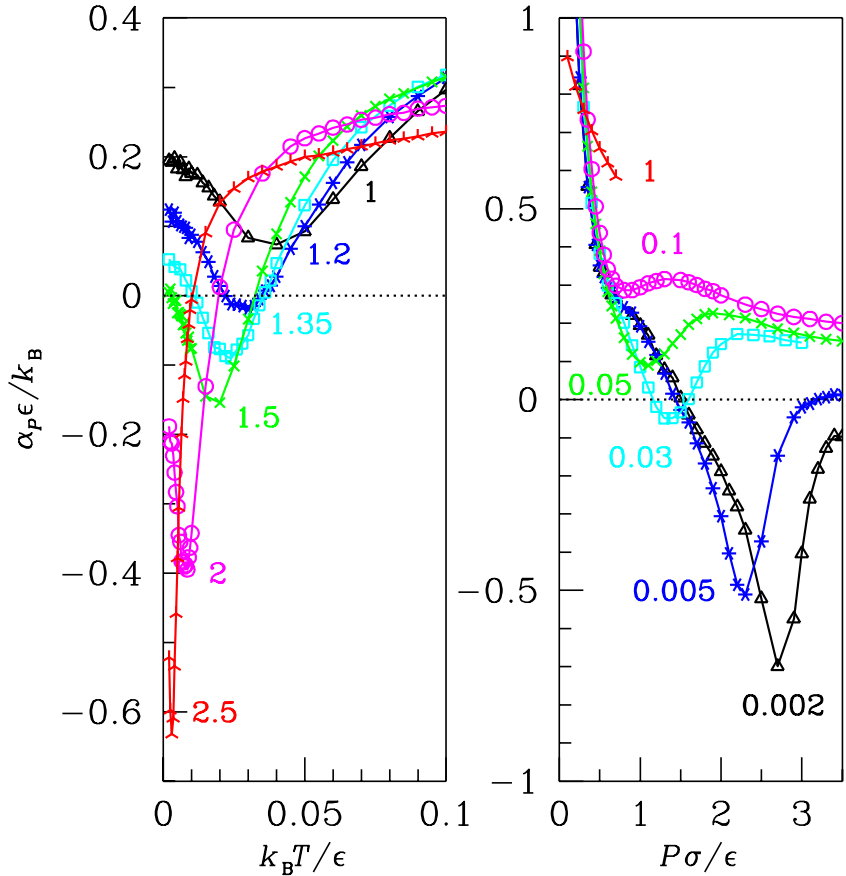}}
\caption{Thermal expansion coefficient plotted as a function of the reduced temperature for different pressures (left panel) and of the pressure for different temperatures (right panel); left panel: same legends as in Fig.~\ref{density}; right panel: triangles (black), $\tau=0.002$; stars (blue), $\tau=0.005$; squares (cianide), $\tau=0.03$; crosses (green), $\tau=0.05$; circles (magenta), $\tau=0.1$; tripods (red), $\tau=1$.}
\label{alfa}
\end{figure}
\begin{figure}
\centerline{\includegraphics[width=0.75\textwidth]{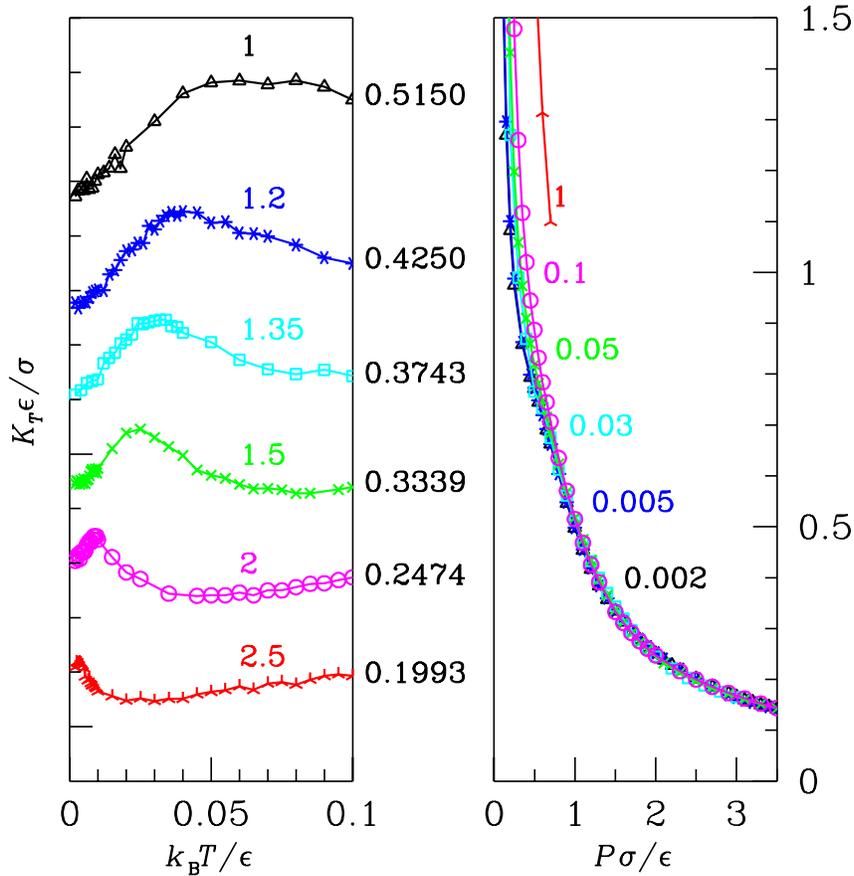}}
\caption{Isothermal compressibility plotted as a function of the reduced temperature for different pressures (left panel) and of the pressure for different temperatures (right panel); same legends as in Fig.~\ref{alfa}. To improve readability, the curves in the left panel were shifted vertically by changing amounts; the separation between two major tick marks is $0.05$, while the absolute scale can be fixed through the value of each curve at $\tau=0.1$, that is reported along the right vertical axis.}
\label{Kt}
\end{figure}
\begin{figure}
\centerline{\includegraphics[width=0.75\textwidth]{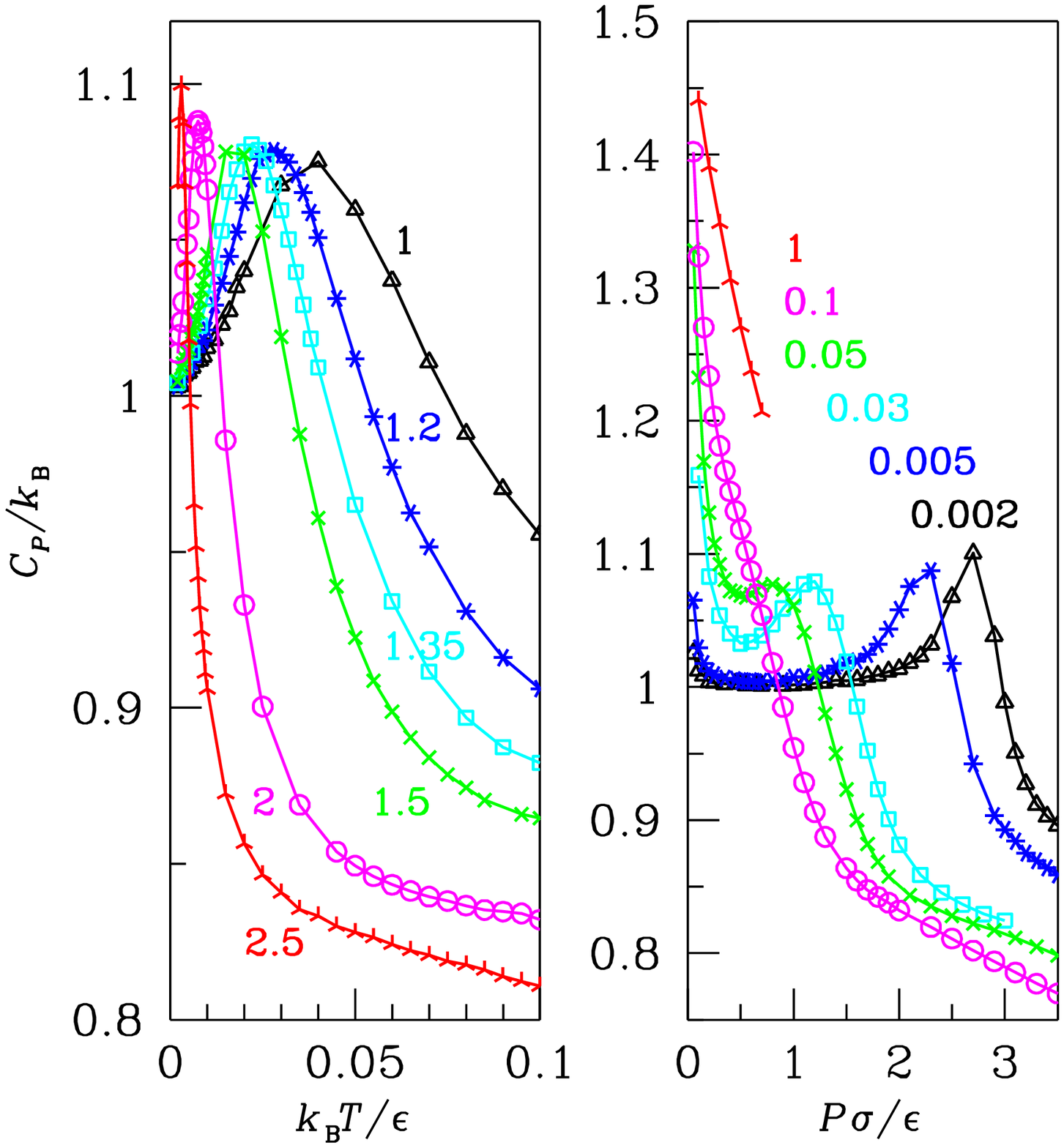}}
\caption{Isobaric specific heat plotted as a function of the reduced temperature for different pressures (left panel) and of the pressure for different temperatures (right panel); same legends as in Fig.~\ref{alfa}.}
\label{Cp}
\end{figure}
\begin{figure}
\centerline{\includegraphics[width=0.75\textwidth]{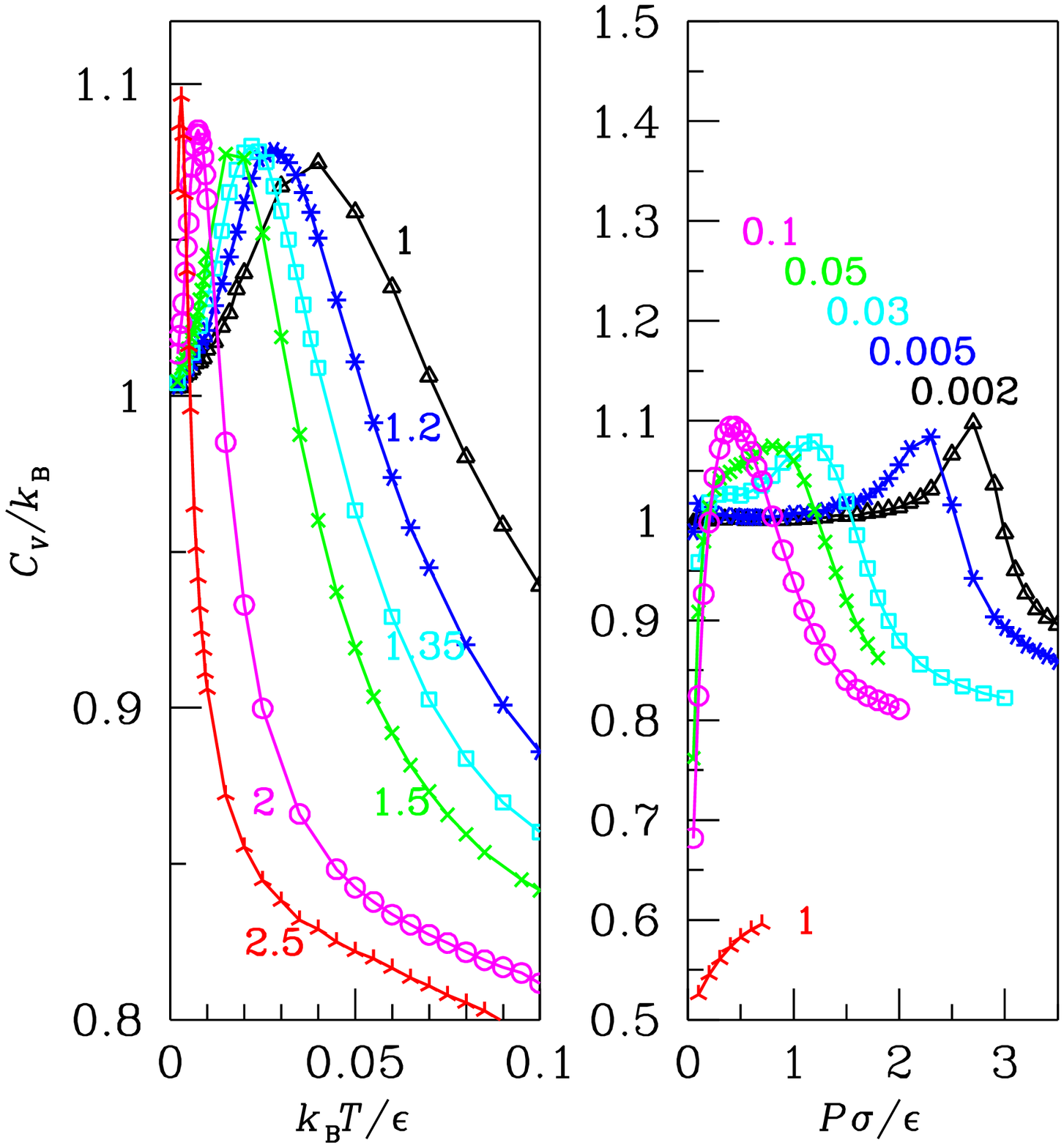}}
\caption{Isochoric specific heat plotted as a function of the reduced temperature for different pressures (left panel) and of the pressure for different temperatures (right panel); same legends as in Fig.~\ref{alfa}.}
\label{Cv}
\end{figure}

The temperature $T_{\mathrm{M}}$ is the highest temperature at which the fluid manifests the volumetric anomaly, associated with the emergence of a single or double extremum in the density. This anomaly obviously reverberates in the thermodynamic response functions. Rolle's theorem applied to an isothermal or isobaric ``cut'' of the anomalous region, producing two distinct intersections with the density extrema lines where $\alpha_P=0$, implies that the expansivity must have a minimum at some point in between. The isobaric minimum of $\alpha_P$ as a function of $T$ (see Fig.~\ref{alfa}) goes along (though not exactly coinciding) with an inflection point of the density; both such features were found to survive well outside the anomalous region, down to $\Pi\simeq0.6$ (see also Fig.~\ref{density}). The minimum-expansivity locus radiates out of this region at the point C of confluence of the TMD and TmD lines (see Fig.~\ref{loci}).

Sastry and coworkers~\cite{sastry} have showed that the isothermal compressibility {\em increases on cooling} at any point along the TMD line wherever its slope is negative and the second temperature derivative of the volume, evaluated at the same point, is positive $\left[(\partial^2v/\partial T^2){_{P,\,\mathrm{at\,TMD}}}=v(\partial\alpha_P/\partial T){_{P,\,\mathrm{at\,TMD}}} > 0 \right]$. The above conditions apply for $P>P_{\mathrm{M}}$. A similar thermodynamic constraint implies that the isothermal compressibility {\em increases on heating} at any point along the TmD line where the slope is negative and such also is the second temperature derivative of the volume, which is the case of the 1D GCM fluid. The necessary consequence of this twofold constraint is the emergence of a maximum in the isothermal compressibility at some temperature in the interval $T_{\mathrm{min}}(P)<T<T_{\mathrm{max}}(P)$. As shown in Figs.~\ref{loci} and \ref{Kt}, the maximum persists even for $\Pi > 1.5$, when the minimum of the density has disappeared. We can see that, as originally predicted by Sastry and coworkers~\cite{sastry}, the locus of the compressibility maxima intersects the TMD line exactly at its maximum M, where $(dT{_{\mathrm{max}}(P)/dP)_{P_{\mathrm M}}}=0$.

Thermodynamic consistency implies~\cite{sastry} that the anomalous temperature dependence of $K_T$ should also reflect in the behaviour of $C_P$ as a function of $T$ at fixed $P$ (see the left panel of Fig.~\ref{Cp}). As the temperature drops to zero, the isochoric specific heat approaches unity ({\em i.e.}, the value of an interacting 1D fluid at $T=0$), as also does the isobaric specific heat since $C_P=C_V+Tv\alpha_P^2/K_T$ (see Fig.~\ref{Cv}); $C_P$ shows then a maximum, whose $(P,T)$ locus runs very close to the isobaric minimum-expansivity line (see Fig.~\ref{loci}). We also note that the height of the $C_P$ maximum changes very little with either the pressure or the temperature, an aspect that we shall return to soon.

Upon heating the system, the thermodynamic properties of the fluid should asymptotically approach their ideal-gas values. As a result, in order for their low-temperature behaviour to match the high-temperature regime, a second extremum must develop in all of the three response functions, besides the one at low temperature that we have already commented on: this second extremum is a maximum in $\alpha_P$ which, as $T \rightarrow \infty$, must tend to zero as $1/T$; instead, a minimum is found in both $K_T$ and  $C_P$ which should approach $1/P$ and $3/2$, respectively. Hence, the second extremum exhibited by the response functions is the necessary outcome of the existence of two different limiting behaviours and, as such, does not arise from any additional independent anomaly of the fluid at high temperature.

The line of the isobaric temperature minima of $K_T$ has also been traced in Fig.~\ref{loci}, where it is apparent that this locus and the locus of the isobaric temperature maxima of $K_T$ stem out of the same point, located at $\Pi_{K_T}\simeq0.84$ and $\tau_{K_T}\simeq0.11$. These two lines get closer again at high pressure.

The right panels of Figs.~\ref{alfa}, \ref{Kt}, \ref{Cp}, and \ref{Cv} show how the fluid reacts to an isothermal compression. Both the density and the energy were found to steadily increase with increasing pressure. Moreover, the isothermal compressibility is seen to decrease monotonically with $P$, its value being not significantly affected by the temperature over the explored range. Instead, $\alpha_P$ and $C_P$ exhibit a more complex behaviour. As for the expansivity, the thermodynamic relation $(\partial \alpha_P / \partial P)_T = - (\partial K_T / \partial T)_P$ implies that isobaric temperature extrema of the compressibility should map onto isothermal pressure extrema of the expansivity~\cite{rebelo}. In fact, Fig.~\ref{loci} shows that the isothermal pressure minima (maxima) of $\alpha_P$ do actually fall, to within the numerical uncertainty of the calculations, along the locus of the isobaric temperature maxima (minima) of $K_T$.

A maximum is also present, at low temperature, in the isobaric specific heat as a function of $P$ at constant $T$ (see the right panel of Fig.~\ref{Cp}); as expected, this quantity is $3/2$ at $P=0$ and, correspondingly, $C_V=1/2$. We found that the locus of the pressure extrema of $C_P$ runs on top of the temperature minima locus of the expansivity (see Fig.~\ref{loci}). The thermodynamic relation~\cite{callen}
\be
\left( \frac{\partial C_P}{\partial P} \right)_T = -vT\left[ \alpha_P^2 + \left( \frac{\partial \alpha_P}{\partial T} \right) _P \right]
\label{relation}
\ee
states that, if $\alpha_P=0$, an isobaric temperature extremum of $\alpha_P$ should coincide with an isothermal pressure extremum of $C_P$. This is precisely what happens at the confluence point of the TMD and TmD lines. Elsewhere, the balance between $\alpha_P^2$ and a negative temperature derivative of $\alpha_P$ locates the zero of $(\partial C_P / \partial P)_T$ at lower temperatures with respect to the locus of vanishing $(\partial \alpha_P / \partial T)_P$. However, upon moving inside the anomaly region, the first term on the r.h.s of Eq.~\ref{relation} turns out to be very small (of the order of $10^{-4}$) for pressures just greater than $\Pi_{\mathrm C}$, and becomes even smaller with increasing pressure because of the gradual drop of both the volume and temperature. As a result, the pressure maxima locus of $C_P$ runs very close to the temperature extrema loci of $\alpha_P$ {\em and} of $C_P$ itself (see Fig.~\ref{loci}). This latter circumstance, {\em i.e.}, the nearby coincidence (to within the uncertainty of the present calculations) of the pressure and temperature maxima loci of $C_P$, can be explained by the implicit function theorem:
\be
\left( \frac{\partial X}{\partial P} \right)_T = - \left( \frac{\partial X}{\partial T} \right) _P \cdotp \left( \frac{\partial T}{\partial P}\right)_{X}
\label{ift}
\ee
where $X(P,T)$ is a generic property of the system and the second partial derivative on the r.h.s. of Eq.(\ref{ift}) is evaluated along a constant-$X$ thermodynamic path. In general, whenever $X$ has an extremum as a function of $P$ at given $T$ ({\em i.e.}, $\left( \partial X / \partial P \right)_T = 0$), the derivative $\left( \partial T / \partial P \right)_X$, evaluated along a constant-$X$ path, vanishes as well (a similar argument naturally holds upon inverting $P$ with $T$). This obviously implies that a pressure (temperature) extremum of $X$ is {\em not} necessarily associated with a temperature (pressure) extremum of the same quantity. However, it can be shown that, when $X(T,P)$ displays a ``ridge'' of equal-height extrema, {\em both} the pressure and the temperature derivatives of $X$ vanish there (and {\em vice versa}), which means that the extremum is such along both thermodynamic axes. To a very good approximation, this is actually what happens for the isobaric specific heat of the 1D GCM fluid.

At very low temperature the peak of the isobaric specific heat is preceded by a broad minimum, actually almost a plateau at one. This latter circumstance is explained by the very small values attained by the quantity $vT\alpha_P^2/K_T$, which yields the difference between $C_P$ and $C_V$; moreover, $\alpha_P$ vanishes at a point located within the implicated pressure interval. In passing, we note that $C_P=C_V$ along the TMD and TmD lines.

One more aspect of this thermodynamic scenario still needs to be outlined. The volumetric anomaly is accompanied by an anomaly of the entropy in that, whenever $\alpha_P<0$, the entropy {\em increases} with $P$ at constant temperature since $(\partial S/\partial P)_T=-(\partial V/\partial T)_P$. In order to illustrate this aspect, we calculated the total entropy per particle of the fluid through the Euler relation, $s=\beta(u+Pv-\mu)$. As for the chemical potential, given its value at some reference state, one can calculate its value at any other state upon integrating a suitable thermodynamic property along either an isothermal or isobaric path, under the condition that no coexistence locus is being crossed along the path:
\be
\mu(T,P_2)=\mu(T,P_1)-\int_{P_1}^{P_2}{\rm d}P\,v(T,P)\,,
\label{mu1}
\ee
\be
\frac{\mu(T_2,P)}{T_2}=\frac{\mu(T_1,P)}{T_1}-
\int_{T_1}^{T_2}{\rm d}T\,\left[ \frac{u(T,P)+Pv(T,P)}{T^2} \right]\,.
\label{mu2}
\ee
Equations~\ref{mu1} and \ref{mu2} readily follow from the Euler and Gibbs-Duhem relations. The chemical potential at the reference thermodynamic state (a dilute-fluid state) can be estimated using Widom's particle-insertion method~\cite{Widom}. Upon performing a NVT simulation for $\rho=0.2$ and $\tau=0.1$, we obtained the value $\beta\mu=-0.7341$ at a reduced pressure of $0.0296$. Using this value as a parameter in a NPT simulation, carried out at the same temperature as in the constant-volume simulation, we obtained the value $\beta\mu=-0.7348$, which coincides with the corresponding NVT estimate to within the statistical error of the calculation.
\begin{figure}
\centerline{\includegraphics[width=0.7\textwidth]{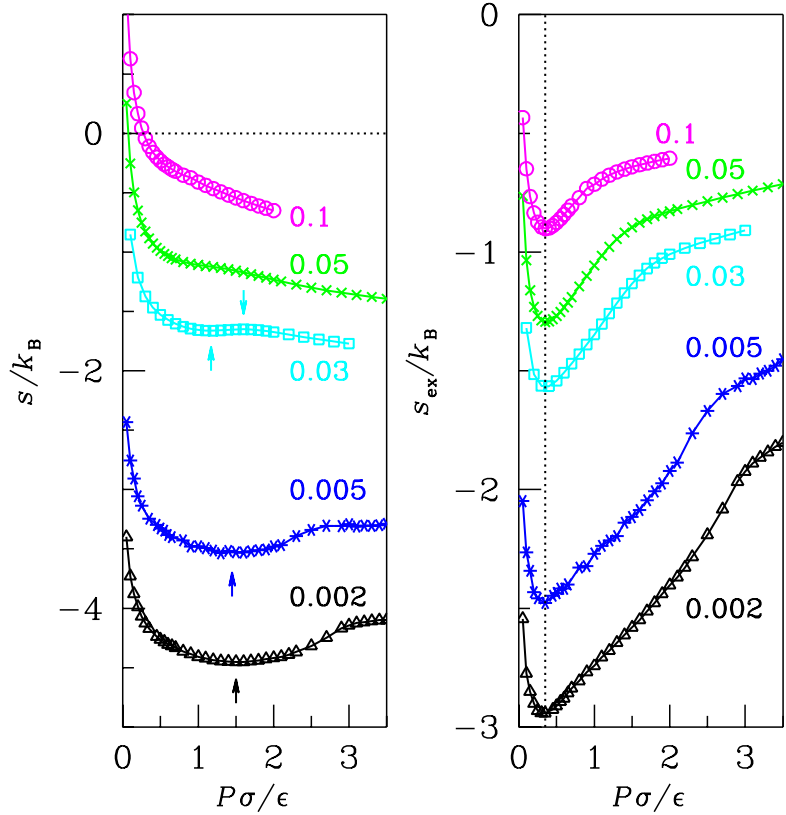}}
\caption{Total entropy (left panel) and excess entropy (right panel) per particle plotted as a function of the pressure for different reduced temperatures: triangles (black), $\tau=0.002$; stars (blue), $\tau=0.005$; squares (cianide), $\tau=0.03$; crosses (green), $\tau=0.05$; circles (magenta), $\tau=0.1$. The total entropy reported on the left panel does not include the additive constant ${\ln}\left[2\pi m/(\epsilon h^2)\right]^{1/2}$; the arrows indicate the positions of the extrema; a dotted line has been traced for $\Pi=0.35$ in the right panel.}
\label{entropy}
\end{figure}

The left panel of Fig.~\ref{entropy} shows the results for the total entropy per particle of the 1D GCM fluid. As expected, at high temperature this quantity decreases with $P$, because of the increasing strength of the positional correlations, a process which typically leads to a reduction of the number of microstates available to the system in a given macrostate. However, as soon as the fluid enters the anomalous region, {\em viz.}, when the temperature drops below $T_{\mathrm M}$, the total entropy develops a minimum that is followed by a maximum at larger pressure, a state beyond which the entropy starts decreasing again. This counterintuitive behaviour is a direct consequence of the volumetric anomaly; correspondingly, the locations of the two pressure extrema change with the temperature, following the TMD and TmD lines.

The right panel of Fig.~\ref{entropy} shows that a minimum is also present, at lower pressure, in the excess entropy of the fluid, $S_{\mathrm{ex}}=S-S_{\mathrm{id}}$, where $S_{\mathrm{id}}$ is the entropy of an ideal gas having the same density and temperature of the GCM fluid. However, three notable differences show up in the thermodynamic behaviours of the excess and total entropy, respectively: i) the minimum of the excess entropy is located at $\Pi_{\mathrm{min}}^{(s_{\mathrm{ex}})}\simeq 0.35$ and its position does not appreciably shift with $T$ over the explored range; ii) for $\Pi>\Pi_{\mathrm{min}}^{(s_{\mathrm{ex}})}$ the excess entropy rises to zero and does not exhibit any other extremum; iii) the minimum persists in $s_{\mathrm{ex}}$ even for $T>T_{\mathrm M}$. All such circumstances suggest that the pressure threshold at $\Pi_{\mathrm{min}}^{(s_{\mathrm{ex}})}$ may be a significant indication of the existence of two markedly different structural regimes in the fluid. We shall come back to this point in the following section. However, before concluding this outline of the thermodynamic properties of the model, we note that the excess entropy of the GCM fluid displays a minimum also in upper dimensions (2D, 3D). For temperatures slightly above the maximum melting temperature, this minimum occurs at a reduced density corresponding to an average interparticle distance of approximately $1.53\sigma$, a value that is fairly congruent with the values in 1D, which range between $1.46\sigma$ and $1.56\sigma$ as the reduced temperature rises from $0.002$ to $0.1$. In 2D and 3D the excess-entropy minimum falls just beyond the density corresponding to the maximum melting temperature, which unambiguously marks the border between two different thermodynamic conditions: a low-density regime in which the system behaves as a normal fluid and, upon compression, eventually freezes; a high-pressure regime in which, instead, the crystal re-melts, with increasing pressure, into a fluid whose properties are very different from the ordinary ones. 
\subsection{Structural properties}
\begin{figure}
\centerline{\includegraphics[width=0.75\textwidth]{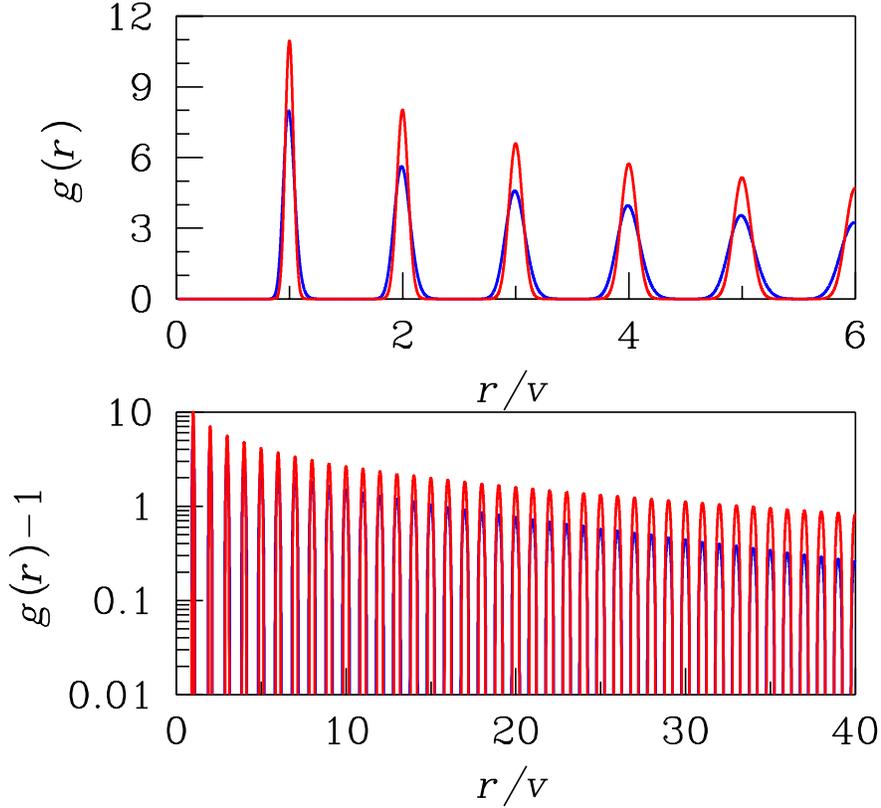}}
\caption{Radial distribution function at $\tau=0.002$ plotted as a function of the distance between particles, relative to their average separation at the corresponding number density, for $\Pi=0.05$ (blue) and $\Pi=0.35$ (red); the lower panel shows the decay of the total correlation function at larger distances on a semilogarithmic scale.}
\label{rdf1}
\end{figure}

In this section we shall focus on the structural properties of the 1D GCM at low temperature $(\tau=0.002)$. Upon isothermally compressing the initially dilute fluid, the local order is more and more enhanced, and also extends to larger and larger distances. As seen in Fig.~\ref{rdf1}, the RDF already looks highly structured at a relatively low pressure ($\Pi=0.05,\rho=0.47$), showing a quasi-crystalline profile characterized by a series of sharply defined coordination shells (each subtending an integrated conditional number density of one particle), with period equal to the average interparticle distance and no appreciable overlap between adjacent shells up to fairly large distances. Such a regularly equispaced arrangement is analogous to that exhibited by a 1D gas of hard particles for packing fractions approximately larger than $83\%$~\cite{pvg}, a value corresponding to a reduced density of the present model equal to $0.33$ at a reduced temperature of $0.002$. In this respect, the core-softened GCM fluid matches and extends to higher densities the quasi-crystalline phase behaviour observed in the ``isomorphic'' hard-core Tonks gas approaching close packing.
\begin{figure}
\centerline{\includegraphics[width=0.75\textwidth]{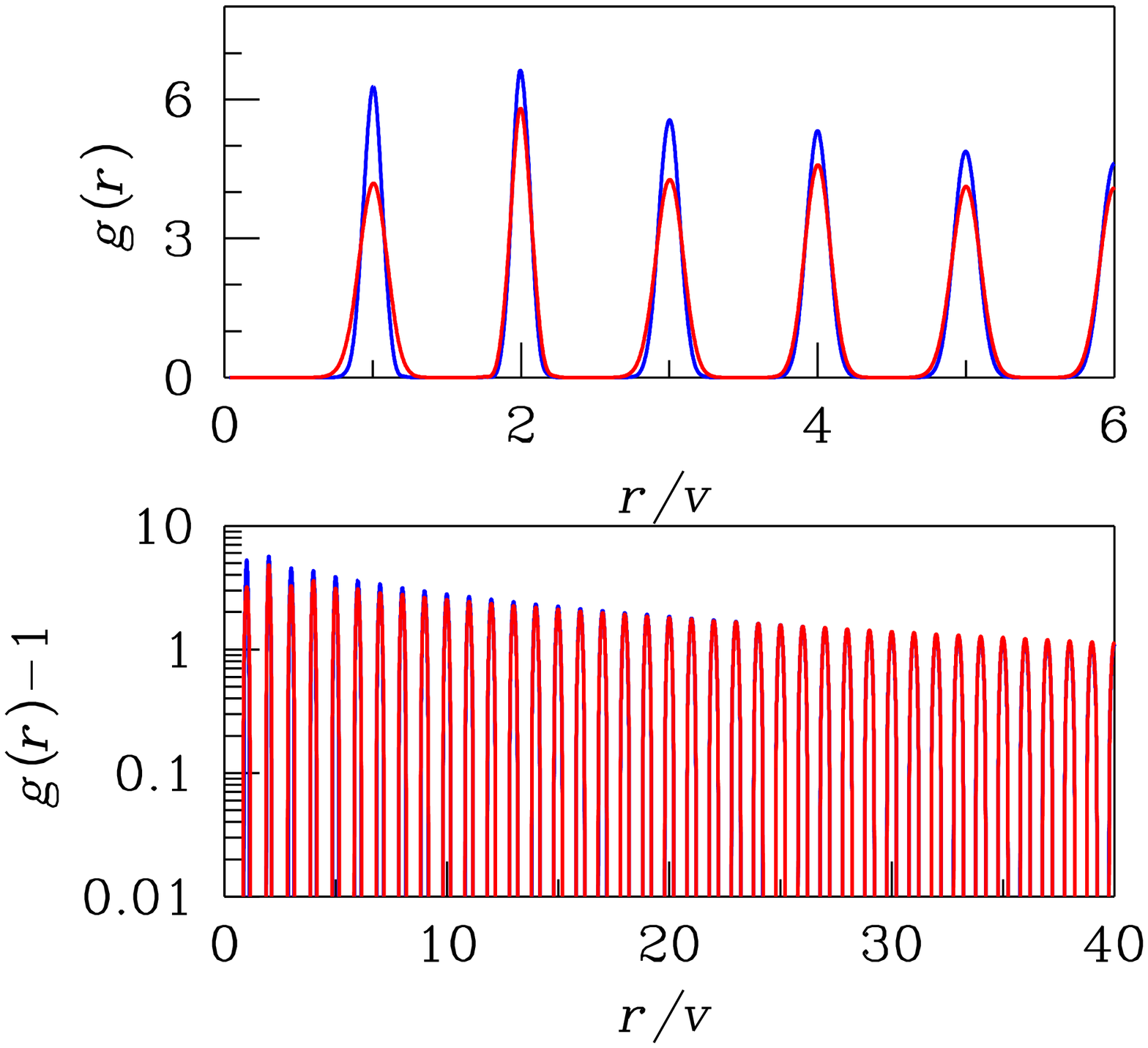}}
\caption{Radial distribution function at $\tau=0.002$ plotted as a function of the distance between particles, relative to their average separation at the corresponding number density, for $\Pi=1.5$ (blue) and $\Pi=2$ (red); the lower panel shows the decay of the total correlation function at larger distances on a semilogarithmic scale.}
\label{rdf2}
\end{figure}
\begin{figure}
\centerline{\includegraphics[width=0.75\textwidth]{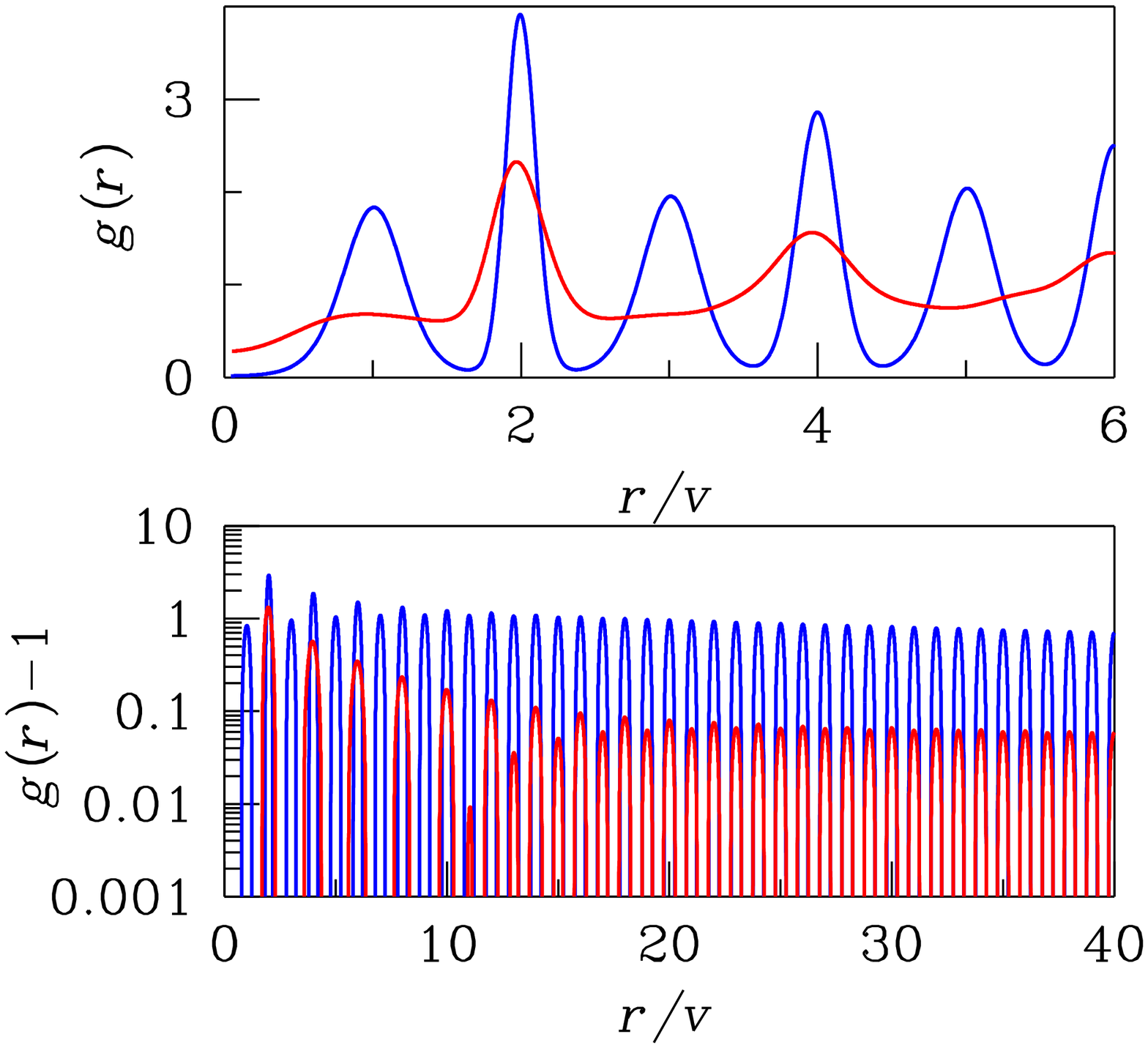}}
\caption{Radial distribution function at $\tau=0.002$ plotted as a function of the distance between particles, relative to their average separation at the corresponding number density, for $\Pi=2.7$ (blue) and $\Pi=3.5$ (red); the lower panel shows the decay of the total correlation function at larger distances on a semilogarithmic scale.}
\label{rdf3}
\end{figure}
\begin{figure}
\centerline{\includegraphics[width=0.75\textwidth]{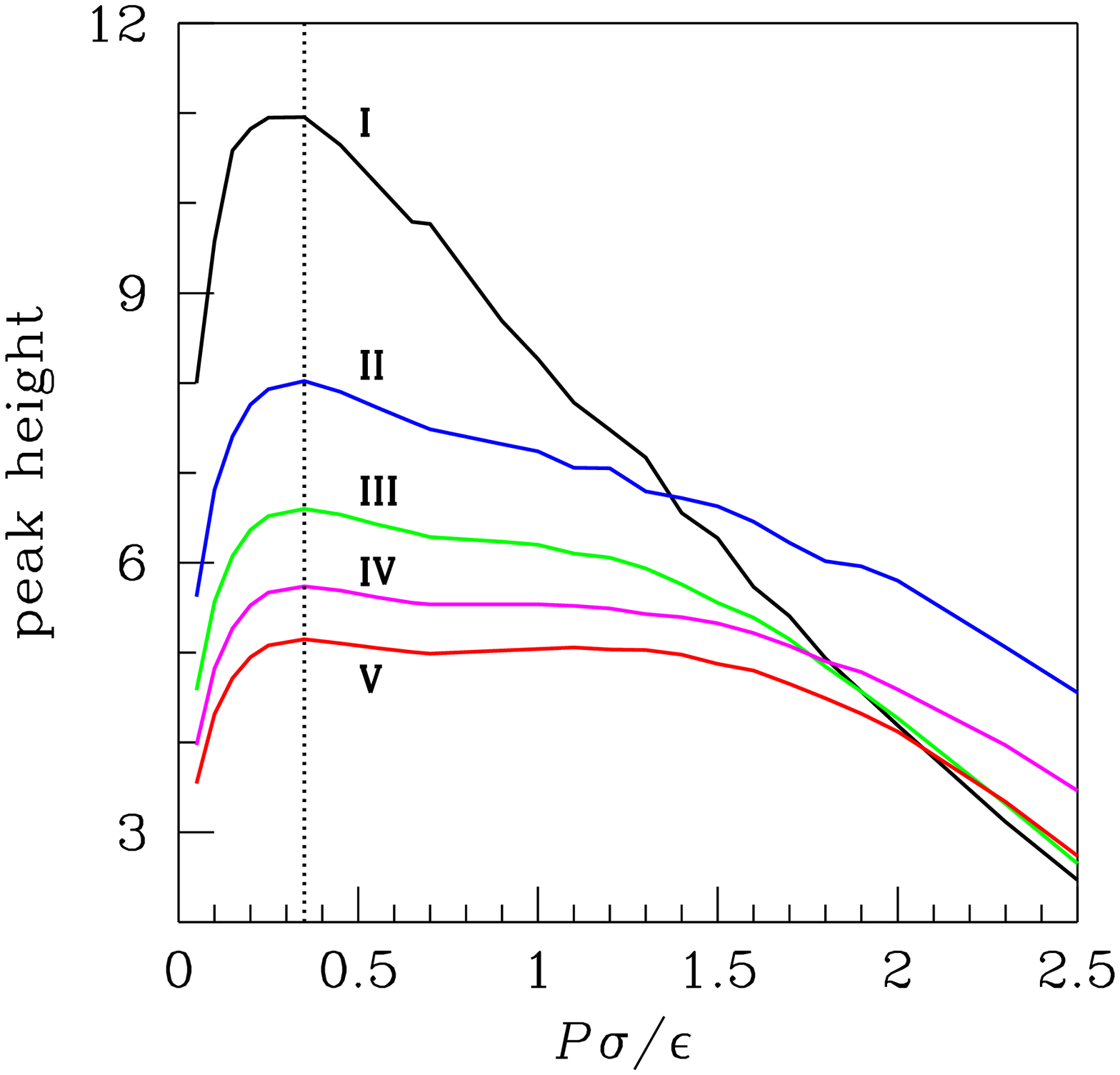}}
\caption{Heights of the first five maxima of the RDF, identified with Roman numerals, plotted as a function of the reduced pressure at $\tau=0.002$.}
\label{maxima}
\end{figure}

As expected for an ordinary simple fluid, the peaks of the RDF get, with increasing pressure, sharper and taller; contextually, the size ($R_0$) of the domain over which the minima between adjacent peaks have, to practical effects, dropped to zero also expands with $P$. Such a trend comes to a stop and is eventually reversed when the pressure increases to values close to the minimum threshold discussed above for the excess entropy (see Figs.~\ref{rdf2} and \ref{rdf3}). In fact, for $\Pi\simeq 0.30$, both the height of the first RDF peak as well as $R_0$ attain a maximum (of about $11$ and $20v$, respectively) and start decreasing thereon with $P$. Figure~\ref{maxima} shows that a similar inversion of the low-density trend is also observed in the other maxima of the RDF at reduced pressures close to $0.35$. Hence, this threshold corresponds to the maximum growth and expansion of the local (hard-core-like) crystalline order in the fluid: in fact, for larger pressures, we observe a global attenuation of the density correlations which reduces the entropic distance of the fluid from its ideal-gas counterpart.

The modifications of the local density profile outlined above prelude the emergence, promoted by the pressure, of a different spatial organization of the fluid. Figure~\ref{maxima} shows that the larger the distance of a given coordination shell from the particle sitting at the origin, the slower the decay of the corresponding peak height with increasing pressure. As a result, the second maximum in $g(r)$ eventually overcomes the first peak for reduced pressures approximately larger than $1.4$, {\em i.e.}, as soon as the fluid enters the volumetric anomaly region. An analogous crossover between the fourth and third peak takes place for $\Pi\simeq 1.75$. A novel type of local order gradually builds up in the fluid, which, for $\Pi\gtrsim 2.5$, shows up in the RDF as the superposition of two distinct modulations, sustained by the odd- and even-numbered peaks, with spatial periods both equal to $2v$ but with different envelopes. Overall, the structure of the fluid is dampened by the pressure, but not everywhere in the same way: odd-numbered peaks get progressively lower and broader than even peaks; moreover, the heights of the even peaks are seen to decay monotonically with the distance, at variance with those of the odd-numbered peaks which display a maximum at some intermediate distance $R_c$; this distance grows with the pressure and appears to saturate, for $\Pi\gtrsim3$, to a size corresponding to the first $12$ coordination shells. For $r>R_c$, the two trains of oscillations do eventually merge into one single oscillation with period $v$.
\begin{figure}
\centerline{\includegraphics[width=0.75\textwidth]{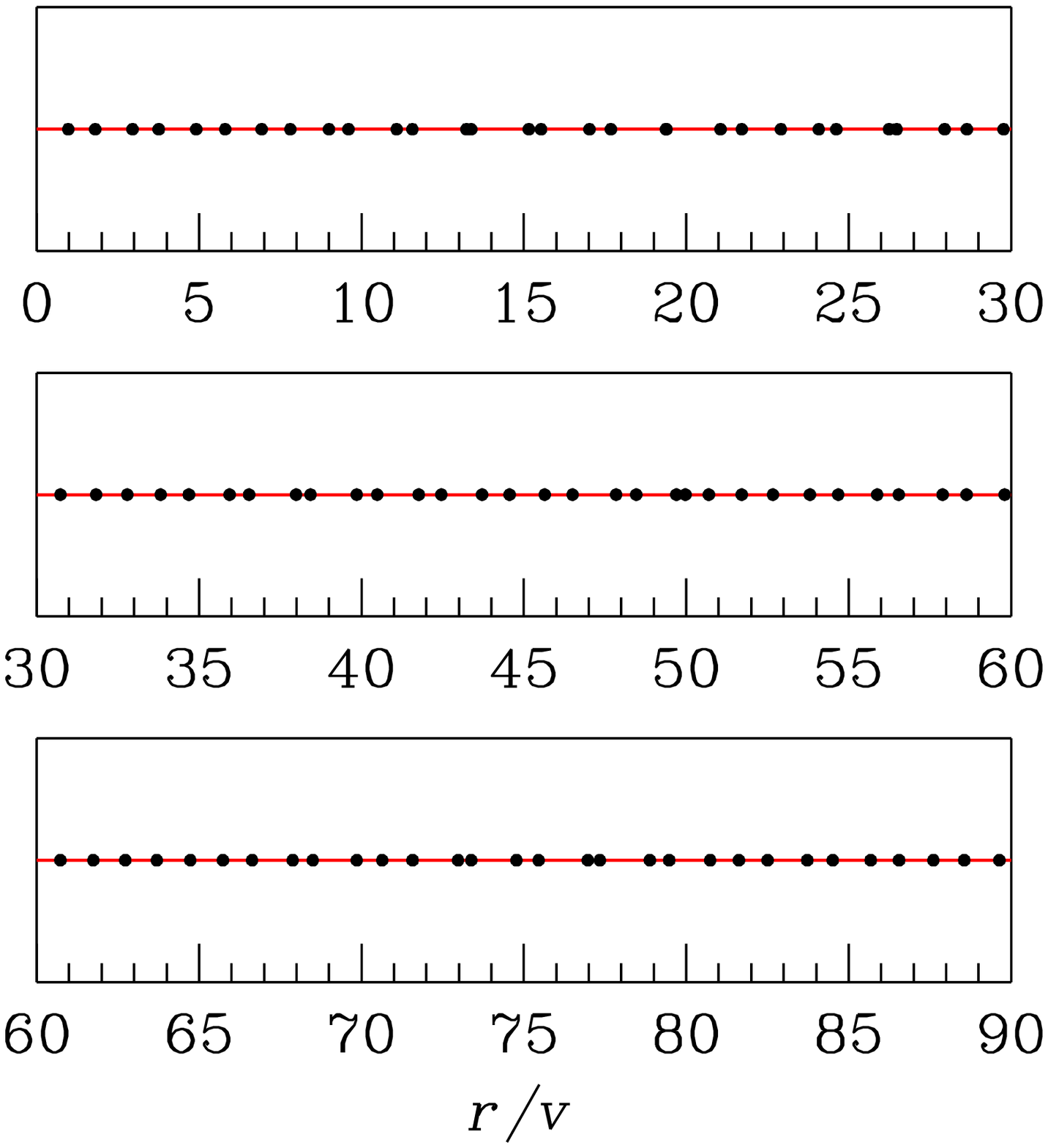}}
\caption{A snapshot of part of the system at $\tau=0.002$ and $\Pi=3~(\rho = 1.8421)$; particles are represented as black dots and distances are relative to the average separation at the given density: we can clearly observe the prevailing presence of regularly spaced patches of ``dimers'' interspersed with ordered sequences of single particles.}
\label{snapshot}
\end{figure}

An indication on the nature of the new spatial organization spawned by the fluid upon compression emerges from a typical snapshot taken at the reduced pressure $\Pi=3$: the presence of patches of particle pairs (``dimers''), regularly spaced with period $2v$ and width of the order of $R_c$, is rather evident in Fig.~\ref{snapshot}. The fact that the even-numbered peaks of the RDF are taller and sharper indicates that the average distance between successive dimers is rather well defined as is the next-nearest-neighbour distance between ``isolated'' particles. On the other hand, the nearest-neighbour coordination shell looks less resolved because, if the particle located at the origin belongs to a dimer, one of its two first neighbours will be closer on average than the other one, belonging to an adjacent dimer. The onset of the pairing phenomenon at high pressure is further witnessed by the value of the RDF at $r=0$, which gives a measure of the fraction of microstates with superimposed particles: this quantity vanishes (to more than six significant figures) for reduced pressures lower than $2.3$ but starts increasing monotonically thereon, being approximately $0.1$ for $\Pi=3$. The crossover from a single-occupancy fluid to a dimer-rich fluid is continuous rather than sharp, {\em i.e.}, it does not entail any thermodynamic transition. We obviously expect that ``trimers'' also form at still higher densities.
\begin{figure}
\centerline{\includegraphics[width=0.75\textwidth]{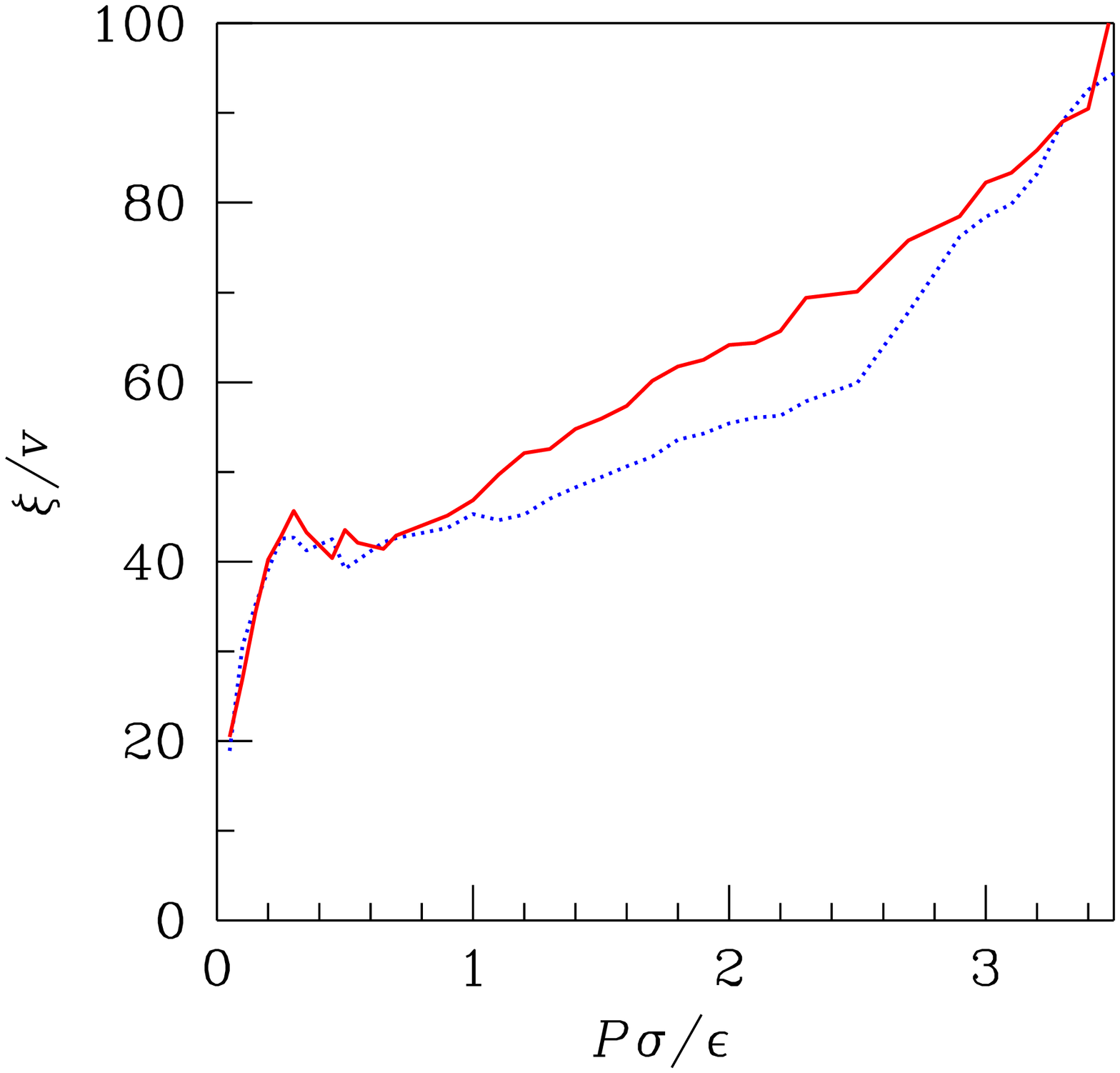}}
\caption{Correlation length, relative to the average separation between particles, plotted as a function of the reduced pressure at $\tau=0.002$: dotted blue curve, $200$ particles; continuous red curve, $500$ particles.}
\label{correlation_length}
\end{figure}
\begin{figure}
\centerline{\includegraphics[width=0.75\textwidth]{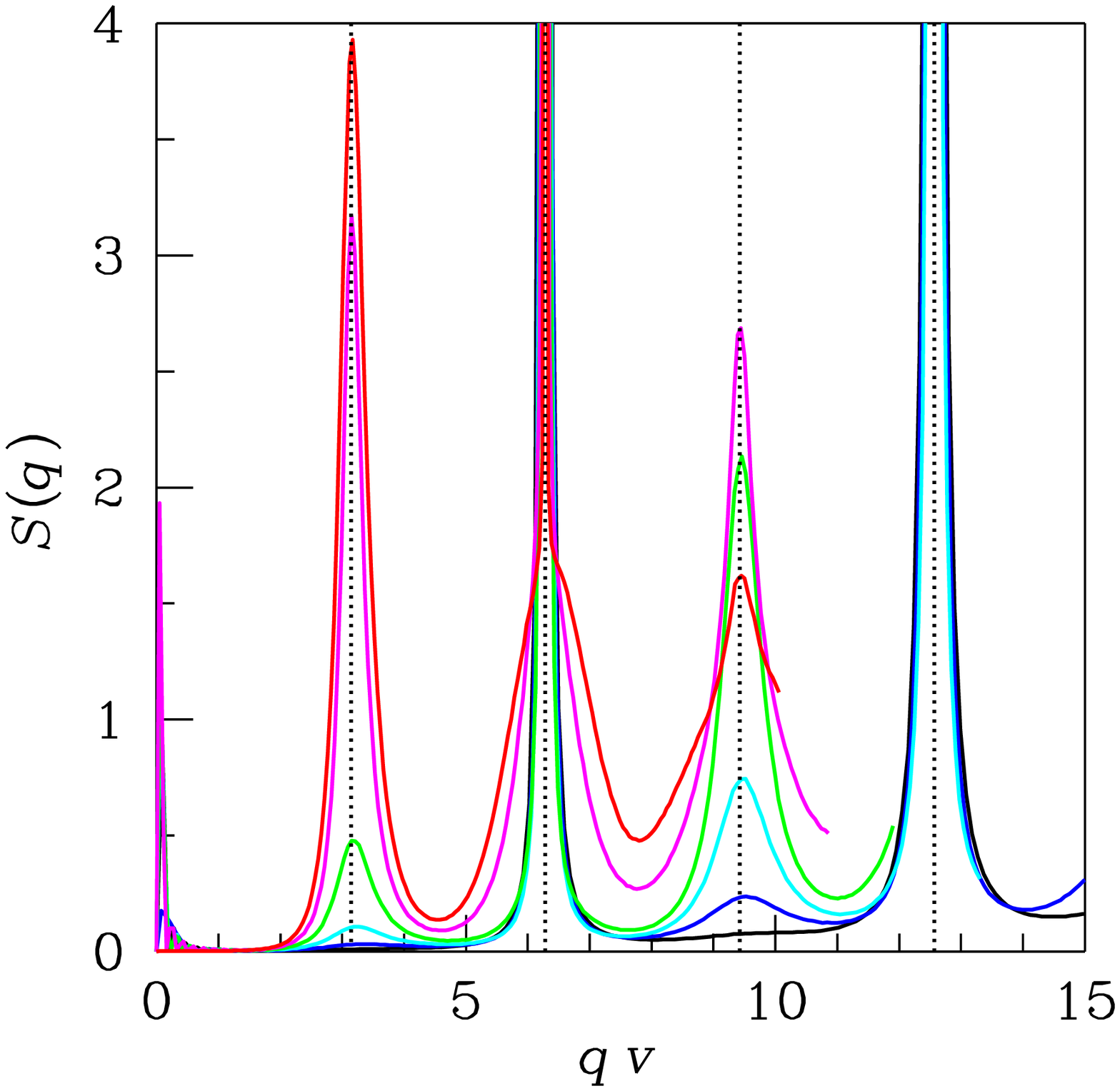}}
\caption{Structure factor plotted as a function of the reduced wavevector at $\tau=0.002$ for $\Pi=$ 1 (black), 1.5 (blue), 2 (cyanide), 2.5 (green), 3 (magenta), 3.5 (red); a specular spectrum is found for negative values of the wavevector. The dotted vertical lines lie in correspondence to the first four multiples of $\pi$. }
\label{S(q)}
\end{figure}

We conclude our discussion of the changes observed in the RDF upon compression with a note on the decay of this function at large distances, which provides an indication on the spatial extent of coherent density fluctuations in the fluid. To this end, we successfully fitted the envelope of more distant maxima to an exponentially decaying function. This very circumstance further confirms the fluid nature of the system even at low temperature and high pressure. The fit allowed us to extract the positional correlation length $\xi$, whose behaviour is shown as a function of $P$ in Fig.~\ref{correlation_length}. When plotted in units of the average separation between particles at a given density, this quantity exhibits a rapid growth for $P\lesssim 0.35$, the structural threshold where the tendency towards a single-occupancy solid-like arrangement produces its strongest effects. For larger pressures, after a slight decrease, the correlation length starts growing again, more slowly than at low pressures, with an approximately linear growth rate for the larger size investigated. We also verified that $\xi$ never exceeds half of the size of the simulated sample, up to the highest pressures investigated.

The structure factor provides a complementary view of the density correlations in the fluid (see Fig.~\ref{S(q)}). At low pressure, the main peaks of $S(q)$ are found at values of the scaled wavenumber $qv\simeq\pm\, 2\pi$, corresponding to a large-distance oscillation of the RDF with period $v$. However, with increasing pressure new satellite peaks -- centred at odd multiples of $\pi$ -- sequentially emerge on the high-wavenumber side, the symmetric peaks at $\pm\,\pi$ being the last to appear (for $\Pi\gtrsim1.4$, again approximately when the fluid enters the anomaly region). Such additional maxima initially concur to modify the profile of the first coordination shell in the RDF, the more so the larger the pressure. However, the increasing relevance of the peaks at $\pm\,\pi$, which eventually overcome in height the peaks at $\pm\,3\pi$ for reduced pressures close to $3$, also signals the emergence of an additional spatial modulation in the local density profile, with period $2v$. 

\section{Concluding remarks}
In this paper we have illustrated the equilibrium properties of a model system of particles interacting, in one dimension, through a bounded repulsive potential with a Gaussian shape. The study of the model was largely carried out using numerical sampling techniques based on the Monte Carlo method. 

As far as the thermodynamic phase stability of the model is concerned, no indication emerged from our analysis of any singular behaviour possibly associated with a fluid-solid or solid-solid transition. However, this very circumstance allowed us to carry out a thorough study of the unexpectedly complex behaviour of the  Gaussian-core model {\em fluid} in one dimension, all the way from high to low temperatures, without incurring any ``break'' caused by phase changes. As a result, we could highlight the full thermal unfolding of a volumetric anomaly -- {\em i.e.}, the increase of the density upon isobarically heating the system -- as a function of pressure, with the emergence of both a maximum {\em and} a minimum of the density with decreasing temperature. While the existence of a maximum in the density of the model had already been observed both in three and two dimensions, evidence of the existence of a minimum, at lower temperature and under appropriate pressure conditions, has never been reported and is being provided here, for the first time, for the one-dimensional fluid. This feature is more unusual than the maximum itself and has been recently observed in metastable states of real substances, most notably in confined supercooled water.

In the light of such findings, we can safely conclude that a soft bounded repulsion like that modelled with a bare Gaussian potential, devoid of any additional repulsive or attractive component, possesses the ``minimal'' requisites that make it possible for the fluid to exhibit waterlike anomalies even in a one-dimensional hosting space. Sadr-Lahijany and coworkers~\cite{scala} had already described the nonstandard thermodynamic properties that emerge from a core-softened potential, explicitly designed to mimic the effect of hydrogen bonding: a repulsion with two length scales set by a hard core and by a finite negative shoulder, followed at larger distances by an attractive well. In this respect, the main result of this paper is that definitely much less is needed to ``turn on'' a volumetric anomaly, even in a one-dimensional fluid: in fact, a {\em finite} softened repulsion, modelled as a {\em one-scale} potential with a downward concavity at short distances and an upward one at larger distances, is sufficient to generate the density anomaly, also bounded by two temperature extrema, as well as a cascade of related peculiarities in all of the three thermodynamic response functions.

Crucial for the thermodynamic onset of the volumetric anomaly in the currently investigated model is the average separation between particles at a given pressure: in fact, we may conjecture that, if such a distance is smaller than the distance $r_0=\sigma/\sqrt{2}$ where the concavity of the Gaussian potential changes from upward to downward, the density will grow upon isobarically heating the fluid even at zero temperature, because the particles will find it (thermodynamically) more advantageous to {\em reduce} their average mutual distance, since this will ultimately entail a significant gain in the entropy of the fluid which overcomes the corresponding increase of internal energy. Such a ``marginal'' condition corresponds to a threshold density $\rho_0=\sqrt{2}$ which, in the ordered ground state formed by equally spaced particles, is attained with a reduced pressure $\Pi_0\simeq 1.77$. Indeed, a similar mechanism is at work in liquid water, in which the breaking of hydrogen bonds, with their hybrid attractive {\em and} repulsive nature~\cite{blandamer}, produces a gradual ``collapse'' of nearest-neighbour molecules inside the inner region of close-contact distances, with a corresponding increase of the density upon heating.

On the other hand, we can also presume that, if the average separation between particles is larger than $r_0$, the density of the fluid will initially decrease upon heating the system from a state at $T=0$ (because the energy decreases as well) until the particles acquire enough kinetic energy to sample the inner region of the potential and thus discover that they may exploit there a more favourable condition, leading to an inversion of the density trend as a function of $T$. On this basis, we may then expect that for $\Pi<\Pi_0$, and for not too low pressures, the anomalous region is bounded from below by a nonzero temperature at which the density shows a minimum.

The investigation of the structural properties of the model also revealed an interesting and unusual scenario. Two structural regimes were distinguished in the equilibrium behaviour of the 1D GCM: for reduced pressures lower than 0.35, the system behaves as a ``normal'' fluid in that the local order is progressively enhanced by an isothermal compression. Instead, upon trespassing the above threshold, a further increase of the pressure induces an overall attenuation of density correlations, an effect associated with the bounded nature of the repulsion. However, the ensuing approach of the system to the thermodynamic condition of an ``infinite-density ideal gas'' is accompanied by the emergence of a new type of local order, with pairs of almost superimposed particles giving rise to extended quasi-crystalline clusters of ``dimers''. This phenomenon is clearly resolved in the radial distribution function as an extra modulation, which is also signalled by the appearance of satellite peaks in the structure factor at relative wavenumbers equal to odd multiples of $\pi$.

\section*{Acknowledgments}
One of the authors (PVG) wishes to thank Professor Luciano Reatto -- to whom this paper is presented as a homage for this special issue in his honour -- for his long-lasting friendship, which has accompanied the author across the years. Luciano has always been a guide and an inspiring example, not only for his wide and acknowledged scientific competences but also for his severe and demanding attitude towards academic research, as well as for his enduring service to the Italian condensed-matter-physics community.


\begin{thebibliography}{99}
\bibitem{channels} X. Xu, B. Lin, B. Cui, A. R. Dinner, and S. A. Rice, {\em J. Chem. Phys.} {\bf 132}, 084902 (2010).
\bibitem{velarde} S. Herrera-Velarde, A. Zamudio-Ojeda, and R. Casta\~neda-Priego, {\em J. Chem. Phys.} {\bf 133}, 114902 (2010).
\bibitem{VanHove} L. van Hove, {\em Physica} {\bf 16}, 137 (1950).
\bibitem{cuesta}J. A. Cuesta and A, S\'anchez, {\em J. Stat. Phys.} {\bf 115}, 869 (2004).
\bibitem{marquest}C. Marquest and T. A. Witten, {\em J. Phys. (France)} {\bf 50}, 1267 (1989).
\bibitem{santos1}A. Malijevsk\'y and A. Santos, {\em J. Chem. Phys.} {\bf 124}, 074508 (2006).
\bibitem{tonks}L. Tonks, {\em Phys. Rev.} {\bf 50}, 955 (1936).
\bibitem{Fantoni} R. Fantoni, {\em J. Stat. Mech.}, P07030 (2010).
\bibitem{santos2}A. Santos, R. Fantoni and A. Giacometti, {\em Phys. Rev. E} {\bf 77}, 051206 (2008). 
\bibitem{Stillinger} F. H. Stillinger, {\em J. Chem. Phys.} {\bf 65}, 3968 (1976).
\bibitem{likos}C. N. Likos, {\em Phys. Reports} {\bf 348}, 267 (2001).
\bibitem{Prestipino1} S. Prestipino, F. Saija, and P. V. Giaquinta, {\em Phys. Rev. E} {\bf 71}, 050102(R) (2005).
\bibitem{Prestipino2} S. Prestipino, F. Saija, and P. V. Giaquinta, {\em Phys. Rev. Lett.} {\bf 106}, 235701 (2011).
\bibitem{pvg} P. V. Giaquinta, {\em Entropy} {\bf 10}, 248 (2008).
\bibitem{debenedetti} P. G. Debenedetti, V. S. Raghavan, and S. S. Borick, {\em J. Phys. Chem. B} {\bf 95}, 4540 (1991).
\bibitem{sw} F. H. Stillinger and T. A. Weber, {\em J. Chem. Phys.} {\bf 68}, 3837 (1978); {\bf 69}, 4322 (1978); {\bf 70}, 1074 (1979). 
\bibitem{Likos} A. Lang, C. N. Likos, M. Watzlawek, and H. L\"owen, {\em J. Phys.: Condens. Matter} {\bf 12}, 5087 (2000). 
\bibitem{tscuchiya1} Y. Tscuchiya, {\em J. Phys.: Condens. Matter} {\bf 3}, 3163 (1991).
\bibitem{tscuchiya2} Y. Tscuchiya, {\em J. Phys.: Condens. Matter} {\bf 4}, 4335 (1992).
\bibitem{liu} D. Liu, Y. Zhang, C.-C. Chen, C.-Y. Mou, P. H. Poole, and S.-H. Chen, {\em Proc. Natl. Acad. Sci. USA} {\bf 104}, 9570 (2007).
\bibitem{faraone} Y. Zhang, A. Faraone, W. K. Kamitakahara, K.-H. Liu, C.-Y. Mou, J. B. Le\~ao, S. Chang, and S.-H. Chen, arXiv:1005.5387v3 [cond-mat.soft].
\bibitem{mallamace} F. Mallamace, P. Baglioni, C. Corsaro, J. Spooren, H. E. Stanley, and S.-H. Chen, {\em Riv. N. Cimento} {\bf 34}, 2011 (2011).
\bibitem{poole} P. H. Poole, I. Saika-Voivod, and F. Sciortino, {\em J. Phys.: Condens. Matter} {\bf 17}, L431 (2005).
\bibitem{angell} A. Angell, {\em Nature Nanotech.} {\bf 2}, 396 (2007).
\bibitem{sastry} S. Sastry, P. G. Debenedetti, F. Sciortino, and H. E. Stanley, {\em Phys. Rev. E} {\bf 53}, 6144 (1996).
\bibitem{rebelo} L. P. N. Rebelo, P. G. Debenedetti, and S. Sastry, {\em J. Chem. Phys.} {\bf 109}, 626 (1998).
\bibitem{callen}H. B. Callen, {\em Thermodynamics and an Introduction to Thermostatistics} (John Wiley and Sons, New York, 1985), second edition. 
\bibitem{Widom} B. Widom, {\em J. Chem. Phys.} {\bf 39}, 2808 (1963).
\bibitem{scala} M. R. Sadr-Lahijany, A. Scala, S. V. Buldyrev, and H.. E. Stanley, {\em Phys. Rev. E} {\bf 60}, 6714 (1999).
\bibitem{blandamer}M. J. Blandamer, J. Burgess, and J. B. F. N. Engberts, {\em Chem. Soc. Rev.} {\bf 14}, 237 (1985); M. J. Blandamer, J. Burgess, and A. W. Hakin, {\em J. Chem. Soc., Faraday Trans. 1} {\bf 83}, 1783 (1987). 
\end{thebibliography}
\end{document}